\documentclass[showpacs,floatfix,epsfig,twocolumn,amsmath,amssymb,preprintnumbers]{revtex4}
\usepackage{graphicx}%
\usepackage{bm}%
\usepackage{epsfig} \hyphenpenalty=1000 \usepackage{epsfig}
\def\PHOBOS{P\kern-.34em \lower.4ex\hbox{H}\kern-.12em
\lower.6ex\hbox{O}\kern-.08em \lower-.18ex\hbox{B}\kern-.12em
\lower-.45ex\hbox{O}\kern-.12em \lower.12ex\hbox{S}\ }

\def\sqrtsnn{${\rm \sqrt{s_{_{NN}}}}$}
\def\lnsnn{${\rm ln(\sqrt{s_{_{NN}}})}$}
\def\AuAu{${\rm Au+Au}$}
\def\CuCu{${\rm Cu+Cu}$}

\def\p{${\rm p(\bar{p}) + p}$}

\def\dAu{${\rm d}+{\rm Au}$}

\def\avgNp{${\rm \langle N_{part} \rangle}$}

\def\Np{${\rm  N_{part} }$}
\def\Nsp{${\rm  N_{spec} }$}
\def\Npavg{${\rm \langle N_{part} \rangle}$}

\def\dnchfrag{${\rm dN_{ch}/d\eta'/\langle N_{part}/2 \rangle}$}
\def\dnchNp{${\rm dN_{ch}/d\eta/\langle N_{part}/2 \rangle}$}

\def\dnch{${\rm dN_{ch}/d\eta}$}
\def\Nch{${\rm N^{Tot}_{ch} }$}

\def\avgNp{${\rm \langle N_{part} \rangle}$}

\begin{document}
\title{System Size, Energy and Centrality Dependence of Pseudorapidity
Distributions of Charged Particles in Relativistic Heavy Ion Collisions} \author{ 
B.Alver$^4$,
B.B.Back$^1$,
M.D.Baker$^2$,
M.Ballintijn$^4$,
D.S.Barton$^2$,
R.R.Betts$^6$,
R.Bindel$^7$,
W.Busza$^4$,
Z.Chai$^2$,
V.Chetluru$^6$,
E.Garc\'{\i}a$^6$,
T.Gburek$^3$,
K.Gulbrandsen$^4$,
J.Hamblen$^8$,
I.Harnarine$^6$,
C.Henderson$^4$,
D.J.Hofman$^6$,
R.S.Hollis$^6$,
R.Ho\l y\'{n}ski$^3$,
B.Holzman$^2$,
A.Iordanova$^6$,
J.L.Kane$^4$,
P.Kulinich$^4$,
C.M.Kuo$^5$,
W.Li$^4$,
W.T.Lin$^5$,
C.Loizides$^4$,
S.Manly$^8$,
A.C.Mignerey$^7$,
R.Nouicer$^2$,
A.Olszewski$^3$,
R.Pak$^2$,
C.Reed$^4$,
E.Richardson$^7$,
C.Roland$^4$,
G.Roland$^4$,
J.Sagerer$^6$,
I.Sedykh$^2$,
C.E.Smith$^6$,
M.A.Stankiewicz$^2$,
P.Steinberg$^2$,
G.S.F.Stephans$^4$,
A.Sukhanov$^2$,
A.Szostak$^2$,
M.B.Tonjes$^7$,
A.Trzupek$^3$,
G.J.van~Nieuwenhuizen$^4$,
S.S.Vaurynovich$^4$,
R.Verdier$^4$,
G.I.Veres$^4$,
P.Walters$^8$,
E.Wenger$^4$,
D.Willhelm$^7$,
F.L.H.Wolfs$^8$,
B.Wosiek$^3$,
K.Wo\'{z}niak$^3$,
S.Wyngaardt$^2$,
B.Wys\l ouch$^4$
\\
\vspace{3mm}
\small
$^1$~Argonne National Laboratory, Argonne, IL 60439-4843, USA\\
$^2$~Brookhaven National Laboratory, Upton, NY 11973-5000, USA\\
$^3$~Institute of Nuclear Physics PAN, Krak\'{o}w, Poland\\
$^4$~Massachusetts Institute of Technology, Cambridge, MA 02139-4307, USA\\
$^5$~National Central University, Chung-Li, Taiwan\\
$^6$~University of Illinois at Chicago, Chicago, IL 60607-7059, USA\\
$^7$~University of Maryland, College Park, MD 20742, USA\\
$^8$~University of Rochester, Rochester, NY 14627, USA}
\date{\today}
\begin{abstract}
We present the first measurements of the pseudorapidity distribution
 of primary charged particles in Cu+Cu collisions as a function of
 collision centrality and energy, \sqrtsnn\ = 22.4, 62.4 and 200 GeV,
 over a wide range of pseudorapidity, using the PHOBOS detector.
 Making a global comparison of \CuCu\ and \AuAu\ results, we find that
 the total number of produced charged particles and the rough shape
 (height and width) of the pseudorapidity distributions are determined
 by the number of nucleon participants. More detailed studies reveal
 that a more precise matching of the shape of the \CuCu\ and \AuAu\
 pseudorapidity distributions over the full range of pseudorapidity
 occurs for the same \Np/2A value rather than the same \Np\ value.  In
 other words, it is the collision geometry rather than just the number
 of nucleon participants that drives the detailed shape of the
 pseudorapidity distribution and its centrality dependence at RHIC
 energies.
\end{abstract}
 \maketitle
The advent of \CuCu\ collisions from the Relativistic Heavy Ion
Collider (RHIC) at energies similar to those of the earlier \AuAu\
collisions presents a new opportunity to measure the system size
dependence of important observables using different collision
geometries. The \CuCu\ results are expected to provide critical tests
of the parametric dependence of the pseudorapidity density of charged
particles, \dnch, observed previously in \AuAu\ collisions
\cite{Brahms,BackPRLAuAu,BackPRCAuAu62}. They
significantly extend the range of measurements with respect to the
number of participant nucleons, \Np, compared to \AuAu\ and also allow for
a direct comparison at the same \Np.

The observed \dnch\ is a conceptually well-defined quantity that
reflects most effects that contribute to particle production in
heavy-ion collisions. It is sensitive to the initial conditions of the
system, i.e. parton shadowing, and also to the effects of rescattering
and hadronic final-state interactions. In short, the full
distribution of \dnch\ represents a time-integral of particle
production throughout the entire heavy-ion collision.

In this letter, we present the first measurements of the \dnch\ of
primary charged particles over a broad range, ${\rm |\eta| <5.4}$, for
\CuCu\ collisions at a variety of collision centralities. These
measurements were made in the same detector for \sqrtsnn = 22.4, 62.4
and 200 GeV allowing for a reliable systematic study of particle
production as a function of energy.  The \CuCu\ results are compared
to results from \AuAu\ \cite{BackPRLAuAu,BackPRCAuAu62} and d+Au
collisions \cite{BackPRLdAumin,BackPRCdAu} at similar energies
obtained with the same detector. This led us to perform a
comprehensive examination of particle production in \CuCu\ and \AuAu\
collisions for the same number of nucleon participant
pairs and for the same fraction of total cross section in both
systems, and to study the interplay between energy and system size.\par
\begin{figure}[t]
\centering
\epsfig{file=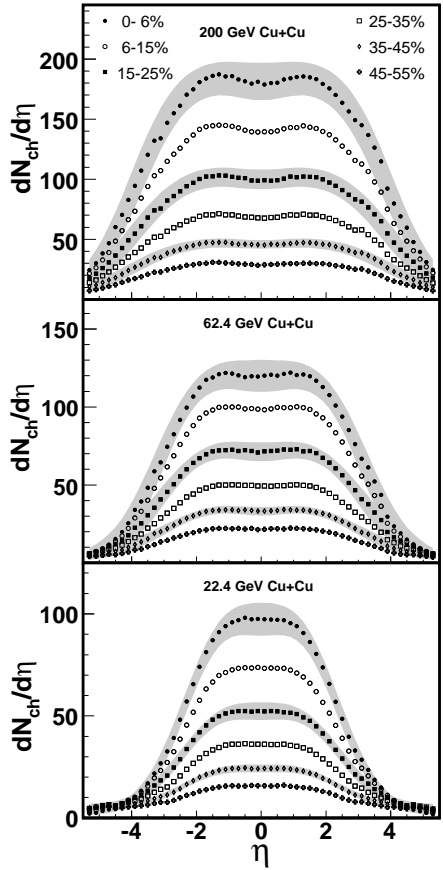,width=6.cm}
\vspace*{-0.4cm}
\caption{\dnch\ distributions of primary charged particles from \CuCu\
  collisions at \sqrtsnn\ = 22.4, 62.4, and 200 GeV for the specified
  centrality bins. The typical systematic errors (90\% C.L.) are shown
  as bands for selected centrality bins. 
\label{fig:fig1}}
\end{figure}
The data were obtained with the multiplicity array of the PHOBOS
detector \cite{BackNim} during the RHIC 2005 run. This array is
identical to that used in our study of\ \AuAu\ and d+Au collisions
\cite{BackPRLAuAu,BackPRCAuAu62,BackPRLdAumin,BackPRCdAu}. Monte Carlo
simulations of the detector performance were based on the HIJING event
generator~\cite{HIJING} and GEANT \cite{GEANT} simulations, folding in
the signal response for scintillator counters (Paddles, covering ${\rm 3.0 <
|\eta| < 4.5}$), and silicon sensors.

The \AuAu\ data at \sqrtsnn = 19.6, 62.4 and 200 GeV used in this
paper for comparison with \CuCu\ results were taken with the PHOBOS
detector \cite{BackPRLAuAu,BackPRCAuAu62}. In the present work, the \AuAu\ data
at 19.6 GeV have been reanalyzed using an improved treatment of dead
detector channels, resulting in slightly smaller multiplicity than
given in previous publications. However, the \dnch\ distributions
shown in the present work and those published in
Refs.~\cite{BackPRLAuAu,BackPRCAuAu62} agree within the systematic
errors. The 19.6 \AuAu\ data and the new \CuCu\ data were
analyzed in the same way as the previously published data of \AuAu\ at
200 and 62.4 GeV \cite{BackPRLAuAu,BackPRCAuAu62}, using two analysis
methods \cite{BackPRL2001}, ``hit-counting'' and ``analog''. In the
present work, the analog results were corrected for the over-counting
of multiply-charged fragments such as He emitted in the forward
regions. The correction, which was largest for the lowest energies and
most peripheral collisions, changed the total number of charged
particles by less than 6\%, and has been taken into account in the
systematic error assignments.
\begin{figure}[t]
\centering
\epsfig{file=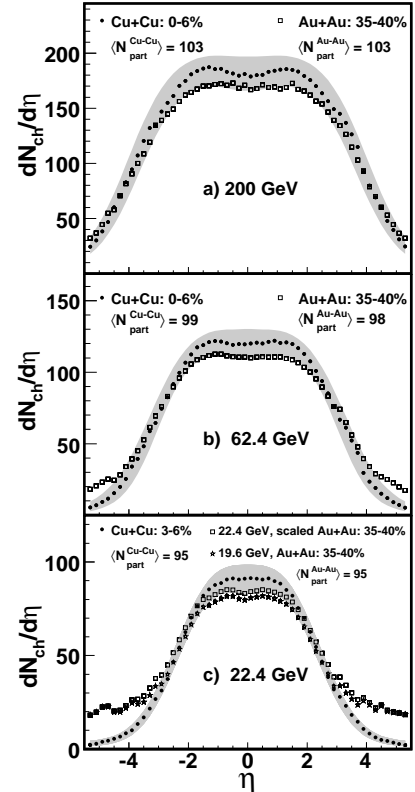,width=5.5cm}
\vspace*{-0.5cm}
\caption{\dnch\ distributions in Cu+Cu and Au+Au
\cite{BackPRLAuAu,BackPRCAuAu62} collisions selected to yield similar
\avgNp, a) at 200 GeV, b) at 62.4 GeV and c) at 22.4 GeV. The grey
band indicates the systematic uncertainty (90\% C.L.) for
Cu+Cu. Errors for \AuAu\ are not shown for clarity.}
\label{fig:fig2}
\end{figure}
\par The centrality determination procedure applied for \CuCu\
collisions is the same as for \AuAu\ collisions at similar energies
\cite{BackPRLAuAu,BackPRCAuAu62}. For the 200 and 62.4 GeV data sets,
the centrality was estimated from the data using the truncated mean of
the Paddles signals. Using several methods, based on the HIJING
\cite{HIJING} and AMPT \cite{AMPT} models, we have estimated our
minimum-bias trigger efficiency for events with a vertex near the
nominal interaction point to be \mbox{$84 \pm 5$\%} and
\mbox{$75 \pm 5$\%} in \CuCu\ collisions at 200 and 62.4 GeV,
respectively. At the lowest energies, 22.4~GeV (\CuCu) and 19.6~GeV
(\AuAu), the much lower beam rapidity ($y_{beam} \sim 3 $) precludes
the use of the Paddles for centrality determination. Instead, we
construct a different quantity, ``EOCT'', the path-length-corrected
sum of the energy deposited in the Octagon (silicon) detector (${\rm |
\eta | \le 3.2}$). This procedure has been discussed in detail in
Ref.~\cite{BackPRLAuAu}. We use a Glauber model calculation
implemented in HIJING simulations to estimate \avgNp\ for each
centrality bin. The \avgNp\ values for various centrality bins for
\CuCu\ collisions are given in Table~\ref{tab:1}.\par
\begin{figure}[t]
\centering \epsfig{file=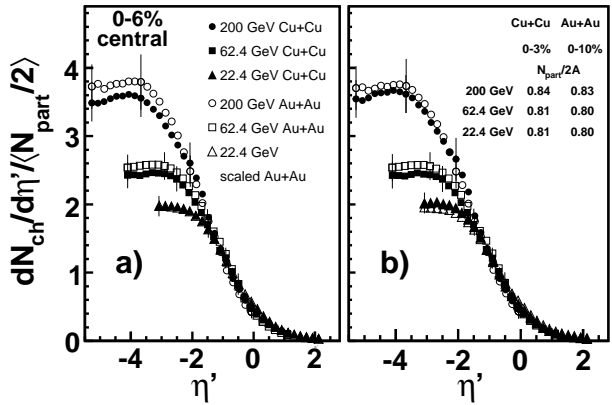,width=8cm}
\vspace*{-0.5cm}
\caption{\CuCu\ and \AuAu\ \cite{BackPRCAuAu62} data at RHIC energies,
plotted as $dN_{ch}/d\eta'/\langle N_{part}/2 \rangle$, where $\eta'
\equiv \eta - y_{beam}$, for a) 0--6\% most central events and b)
events with similar value of \Np/2A. For the 0-6\% central collisions
of \CuCu\ (\AuAu) at 200, 62.4 and 22.4
(19.6) GeV, the values of \Np/2A are 0.82 (0.87),0.78 (0.85) and 0.78
(0.86), respectively. Systematic errors (90\% C.L.)  are shown for
typical points.}
\label{fig:fig3}
\end{figure}
Figure~\ref{fig:fig1} shows the primary charged particle \dnch\
distributions measured in \CuCu\ collisions at \sqrtsnn = 22.4, 62.4,
and 200 GeV for different centrality bins. 
The statistical errors are negligible. Both the height and width of
the \dnch\ distributions increase as a function of energy as has been
seen for \AuAu\ collisions \cite{BackPRLAuAu}.  \par The comparison of
\dnch\ distributions for \CuCu\ and \AuAu\ collisions at the same
energy, for centrality bins chosen so that the average number of
participants in both systems is similar, is presented in
Fig.~\ref{fig:fig2}. No scaling factors are applied. At 200 GeV,
Fig.~\ref{fig:fig2}a, the \dnch\ distributions in both systems at
similar \avgNp\ agree within systematic errors, both in height and
width. At 62.4 and 22.4 GeV, Fig.~\ref{fig:fig2}b and c, we observe
that the distributions agree within systematic errors at midrapidity
but not in the fragmentation regions (i.e. high ${\rm |\eta|}$). The
\dnch\ distributions of \AuAu\ collisions at 19.6 and 62.4 GeV have
been interpolated linearly in \lnsnn\ to obtain the scaled \AuAu\ data
at 22.4 GeV. The fragmentation region dependence on the size of the
colliding nuclei is exhibited through increased charged particle
production in \AuAu\ as compared to \CuCu\ collisions which may be
attributed to the two excited nuclear remnants being bigger in \AuAu\
than in \CuCu\ collisions. This effect is most visible at the lowest
energies where the broad $\eta$ coverage gives access to $|\eta| >
y_{beam}$.  
\par The interplay between the system size and the
collision energy is shown in Fig.~\ref{fig:fig3}. It shows the scaled
pseudorapidity particle densities, \dnchfrag\ (where $\eta' \equiv
\eta - y_{beam}$ corresponds effectively to the rest frame of one of
the colliding nuclei \cite{Brahms,BackPRLAuAu}) for \CuCu\ and \AuAu\
collisions for centrality bins with a) the same fraction of total
cross section 0-6\% and b) similar value of \Np/2A, where A is the
mass number of the colliding nuclei. It should be noted that systems
with matching \Np/2A values will also have matching \Nsp/\Np\ values,
where \Nsp~=~2A~-~\Np\ is the number of non-participating nucleons
(spectators). The results shown in Fig.~\ref{fig:fig3} suggest that in
symmetric nucleus-nucleus collisions the particle density per nucleon
participant pair at the midrapidity region does not depend on the size
of the two colliding nuclei but only on the collision energy and
geometry. In the fragmentation region, the phenomenon of extended
longitudinal scaling observed in \AuAu\ \cite{Brahms,BackPRLAuAu} and
d + Au \cite{BackPRCdAu} collisions is also present in the \CuCu\
data.  The \CuCu\ and \AuAu\ collisions exhibit the same extended
longitudinal scaling in central collisions.  This suggests that the
extended longitudinal scaling holds independent of the collision
energy and system size. The comparison presented in
Fig.~\ref{fig:fig3} reveals an interesting feature: for centrality
bins corresponding to the same fraction of total cross section in
\CuCu\ and \AuAu\ collisions and at similar energy, the \dnchNp\
distributions in both systems agree within errors over the full range
of $\eta$. Slightly better agreement of \dnchNp\ distributions is
obtained for centrality bins selected to yield similar value of \Np/2A
in both systems as presented in Fig.~\ref{fig:fig3}b.
\begin{figure}[t]
\centering
\epsfig{file=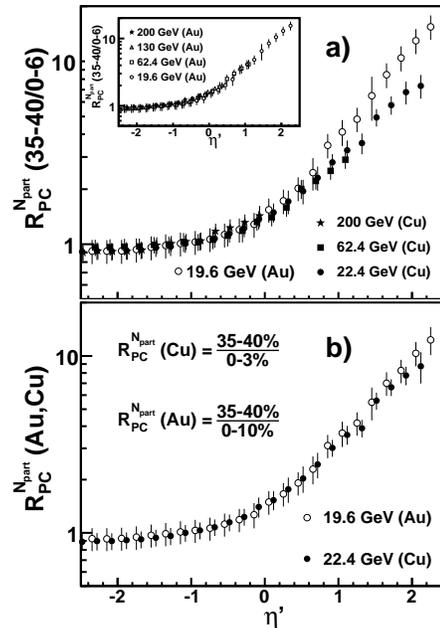,width=6cm}
\vspace*{-0.8cm}
\caption{The ratio, ${\rm R^{N_{part}}_{PC}(\eta')}$, of
$dN_{ch}/d\eta'/\langle N_{part}/2 \rangle$ for \CuCu\ and \AuAu\
collisions at RHIC energies, a) comparing the 35--40\% bin to the
0-6\% most central bin and b) ${\rm R^{N_{part}}_{PC}(\eta')}$ for
\CuCu\ and \AuAu\ for centrality bins selected such that \Np/2A is
similar for the two systems.  The inset figure (a) represents the
${\rm R^{N_{part}}_{PC}(\eta')}$ only for \AuAu\ data for four
different energies \cite{BackPRCAuAu62}. The errors represent a 90\%
C.L.\ systematic error on the ratio. Note : the scaled \AuAu\ at 22.4
GeV has not been added in the figure because the ratio, ${\rm
R^{N_{part}}_{PC}(\eta')}$, for a given system is independent of
energy. The values of \Np/2A are given in Fig.~\ref{fig:fig3} and the
value of \Np/2A for \CuCu\ and \AuAu\ for 35-40\% collisions centrality
at 22.4 (19.6) GeV are both equal to 0.24.}
\label{fig:fig4}
\end{figure}
\begin{figure}[t]
\centering 
\epsfig{file=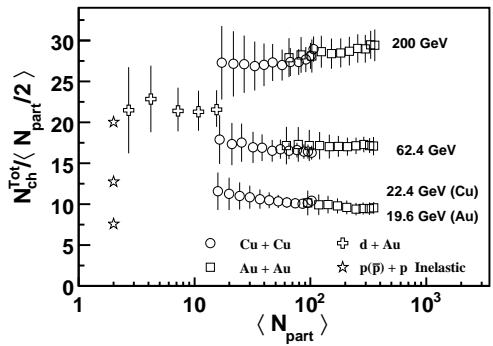,width=6.5cm}
\vspace*{-0.4cm}
\caption{The integrated total of primary charged particle production
obtained by extrapolating the data at each energy into the unmeasured
region is shown as a function of centrality in \CuCu\ collisions. The
\AuAu, d+Au and inelastic ${\rm p(\bar{p})+p}$ data are taken from
Refs. \cite{BackPRCAuAu62,BackPRCdAu}. The uncertainty on \Nch\ and
\Np\ has been included in the error bars.}
\label{fig:fig5}
\end{figure}
\par To study the centrality dependence of extended longitudinal
scaling in \CuCu\ and \AuAu\ collisions, we examine the ratios of
\dnchfrag\ for central and semi-central, denoted ${\rm
R^{N_{part}}_{PC}}$, at different energies as a function of ${\rm
\eta'}$. The inset of Fig.~\ref{fig:fig4}a shows the previously
published results for \AuAu\ collisions \cite{BackPRCAuAu62},
indicating that the change in shape as a function of centrality is
independent of beam energy. The ratios ${\rm R^{N_{part}}_{PC}}$ for
\CuCu\ at three energies exhibit the same feature (solid points) in
Fig.~\ref{fig:fig4}a. By comparing to \AuAu\ results (open points)
this ratio is found to be similar in the midrapidity region ${\rm -2.5
\le \eta' \le 0.5}$ but in the region ${\rm \eta' > 0.5}$ the ratio
for \AuAu\ is higher. To study this difference, we plot in
Fig.~\ref{fig:fig4}b the ${\rm R^{N_{part}}_{PC}}$ ratio for \CuCu\
and \AuAu\ for centrality bins selected to represent similar initial
geometry even more precisely, i.e. similar value of \Np/2A for the two
systems. Using this comparison criterion we observe good agreement
between the two systems over the full range of pseudorapidity.  \par
The values of total charged particle multiplicity, ${\rm
N^{Tot}_{ch}}$, estimated by extrapolation to the unmeasured region of
pseudorapidity, are presented for \CuCu\ collisions at \sqrtsnn\ =
22.4, 62.4 and 200 GeV in Fig.~\ref{fig:fig5} and given in
Table~\ref{tab:1}. The method used to determine ${\rm N^{Tot}_{ch}}$
is detailed in Ref.~\cite{Back2006} for \AuAu\ collisions. At all
energies, ${\rm N^{Tot}_{ch}}$ is obtained by averaging the results of
two techniques. One involved fitting a Wood-Saxon functional form to
the data for $|\eta| \le$ 8 at the two higher energies and ${\rm
|\eta| \le y_{beam}}$ at the lowest energy where ${\rm
y^{CuCu}_{beam}}$(22.4 GeV) = 3.2 and ${\rm y^{AuAu}_{beam}}$(19.6
GeV) = 3.0. The other involved simply integrating the lowest energy
data and using the extended longitudinal scaling result to extrapolate
the higher energy data into the unmeasured regions. The \CuCu\ and
\AuAu\ results are compared at the same energies, 62.4 and 200 GeV, as
well as at nearly the same energy 22.4 (CuCu) and 19.6 GeV (\AuAu) in
Fig.~\ref{fig:fig5}. We observe that \Nch\ scales approximately
linearly with \avgNp\ in both \CuCu\ and \AuAu\ collisions, and has
similar values for the same \Npavg. The comparison indicates that the
transition between inelastic \p\ and \CuCu\ collisions is not
controlled simply by the number of participants, as even the very
central \dAu\ multiplicity per participant pair shows little sign of
continuity to the\ \CuCu\ results.  \par In summary, the measured
pseudorapidity distributions of charged particles and the estimated
total charged particle multiplicity in \CuCu\ collisions are presented
as a function of collision centrality and energy, \sqrtsnn = 22.4,
62.4 and 200 GeV. The results show that \Np\ is the scaling variable
unifying the centrality dependence for \CuCu\ and \AuAu\ in the
midrapidity region and for \Nch. However, the best agreement of the
pseudorapidity distributions per nucleon participants over the full
range of $\eta$ between \CuCu\ and \AuAu\ in the central collisions at the same
energy is obtained for centrality bins selected to yield similar value
of \Np/2A in both systems. The \CuCu\ and \AuAu\ results at similar
energy show that the particle density per nucleon participant pair in
the midrapidity region is similar in both systems. The phenomenon of
extended longitudinal scaling in \CuCu\ and \AuAu\ collisions holds
independent of colliding energy and system size. A dependence on the
size of the colliding nuclei is observed in the pseudorapidity
distributions in the fragmentation region at low energies, 22.4
(\AuAu\ 19.6) and 62.4 GeV, when the collision centrality of the two
systems is selected for similar \Np. This may be attributed to the two
excited nuclear remnants being bigger in \AuAu\ than in \CuCu\
collisions. The essential role of collision geometry when comparing
pseudorapidity distributions of charged particles between nuclear
species is clearly demonstrated.\newline {\small This work was
partially supported by U.S. DOE grants DE-AC02-98CH10886,
DE-FG02-93ER40802, DE-FG02-94ER40818, DE-FG02-94ER40865,
DE-FG02-99ER41099, and DE-AC02-06CH11357, by U.S. NSF grants 9603486,
0245011, 1-P03B-062-27(2004-2007), by NSC of Taiwan Contract NSC
89-2112-M-008-024, and by Hungarian OTKA grant (F049823).}

\begin{table*}[t]
\caption{The estimated number of nucleon participants, \avgNp, and the
total charged particle multiplicity, ${\rm N^{Tot}_{ch}}$,
extrapolated to the unmeasured region for Cu+Cu collisions in different
centrality bins are presented. All errors are systematic (90\% C.L.).
\label{tab:1}
}
\begin{tabular}{c|c|c|c|c|c|c}
\hline
\hline
\multicolumn{1}{ c}{Centrality }&
\multicolumn{2}{|c| }{200 GeV}&
\multicolumn{2}{|c| }{62.4 GeV}&
\multicolumn{2}{c }{22.4 GeV}\\
\hline
Bin (\%)& 
\avgNp &${\rm N^{Tot}_{ch}}$ &
\avgNp &${\rm N^{Tot}_{ch}}$ &
\avgNp &${\rm N^{Tot}_{ch}}$\\ 
       
\hline
0--3 &   106 $\pm$ 3&1541 $\pm$ 70&   
         102 $\pm$ 3&833 $\pm$ 36 &
         103 $\pm$ 3&535 $\pm$ 23 \\       
3--6  &  100 $\pm$ 3&1407 $\pm$ 68  &
         95 $\pm$  3&781 $\pm$ 34  &
         95 $\pm$  3&482 $\pm$ 21  \\
6--10 &    91 $\pm$ 3&1262 $\pm$ 59  & 
           88 $\pm$ 3&721 $\pm$ 32  &
           86 $\pm$ 3&431 $\pm$ 19  \\
10--15 &  79 $\pm$  3&1084 $\pm$ 51  & 
          76 $\pm$  3&635 $\pm$ 27  &
          74 $\pm$  3&375 $\pm$ 18  \\
15--20 &  67 $\pm$ 3&917 $\pm$ 43 & 
          65 $\pm$ 3&541 $\pm$ 24 &
          63 $\pm$ 3&320 $\pm$ 15 \\
20--25 &  57 $\pm$ 3&771 $\pm$ 38  & 
          55 $\pm$ 3&460  $\pm$ 21  &
          53 $\pm$ 3&273  $\pm$ 14  \\
25--30&   47 $\pm$ 3&645 $\pm$ 32 & 
          47 $\pm$ 3&386 $\pm$ 17 &
          44 $\pm$ 3&230 $\pm$ 13 \\
30--35 &  40 $\pm$ 3&538 $\pm$ 27 & 
          38 $\pm$ 3&323 $\pm$15  &
          37 $\pm$ 3&194 $\pm$ 12 \\
35--40&   33 $\pm$ 3&444 $\pm$ 23  & 
          32 $\pm$ 3&270 $\pm$ 13  &
          30 $\pm$ 3&162 $\pm$ 12 \\
40--45 &  27 $\pm$ 3&364 $\pm$ 19 & 
          25 $\pm$ 3&223 $\pm$ 11  &
          25 $\pm$ 3&135 $\pm$ 11 \\
45--50 &  22 $\pm$ 3&293 $\pm$ 15  & 
          21 $\pm$ 3&183 $\pm$ 9 &
          20 $\pm$ 3&112 $\pm$  11 \\
50--55 &  17 $\pm$ 3&234 $\pm$ 13 & 
          16 $\pm$ 3&147 $\pm$ 8 &
          16 $\pm$ 3&92 $\pm$ 11 \\
\hline
\end{tabular}
\end{table*}

\vskip -0.5cm
\begin{thebibliography}{99}
\vskip -1.0cm
\bibitem{Brahms}I. G. Bearden {\it et al.}, Phys.\ Rev.\ Lett.\  {\bf88},202301 (2002).
\bibitem{BackPRLAuAu}B.~B.~Back {\it et al.}, Phys.\ Rev.\ Lett.\  {\bf91}, 052303 (2003).
\bibitem{BackPRCAuAu62}B.~B.~Back {\it et al.}, Phys.\ Rev.\ C {\bf74}, 021901 (2006).
\bibitem{BackPRLdAumin}B.~B. Back {\it et al.}, Phys. Rev. Lett. {\bf93}, 082301 (2004).
\bibitem{BackPRCdAu}B.~B. Back {\it et al.}, Phys.\ Rev.\ C {\bf72}, 031901(R) (2005).
\bibitem{BackNim}B.~B.~Back {\it et al.}, Nucl.\ Instr.\ Meth.\ A {\bf499}, 603 (2003).
\bibitem{HIJING} M.~Gyulassy {\it et al.}, Phys. Rev.\ D {\bf44}, 3501 (1991).
\bibitem{GEANT}GEANT~3.211, CERN Program Library.
\bibitem{BackPRL2001}B.~B.~Back {\it et al.}, Phys.\ Rev.\ Lett. {\bf87}, 102303 (2001).
\bibitem{AMPT}Zi-wei Lin {\it et al.}, Nucl.\ Phys.\ A {\bf698}, 375 (2002).
\bibitem{BackNuclPhys} B.~B.~Back {\it et al.}, Nucl.\ Phys\ A {\bf757}, 28 (2005).
\bibitem{Back2006}B.~B.~Back {\it et al.}, Phys.\ Rev.\ C {\bf74}, 021902(R) (2006).
\end{thebibliography}
\end{document}